\documentclass[journal,10pt,twoside]{IEEEtran}

\usepackage{amsmath,amssymb,bm}
\usepackage{graphicx}
\usepackage{booktabs}
\usepackage{multirow}
\usepackage{algorithm}
\usepackage{algorithmic}
\usepackage{cite}
\usepackage{url}
\usepackage{xcolor}
\usepackage{tabularx}
\usepackage{dblfloatfix}
\usepackage{placeins}
\usepackage{hyperref}

\title{DSTAN-Med: Dual-Channel Spatiotemporal Attention with
Physiological Plausibility Filtering for False Data Injection
Attack Detection in IoT-Based Medical Devices}

\author{
    Md Mehedi Hasan\textsuperscript{a}, Rafiqul Islam\textsuperscript{a}, and Md Zakir Hossain\textsuperscript{b} \\
    \vspace{0.15cm} 
    
    \textsuperscript{a}\textit{School of Computing, Mathematics and Engineering, Charles Sturt University, NSW 2640, Australia} \\
    \textsuperscript{b}\textit{School of Medicine and Psychology, The Australian National University, Canberra ACT 2601, Australia} \\
    \vspace{0.15cm}
    
    \href{mailto:mhasan@csu.edu.au}{mhasan@csu.edu.au}, 
    \href{mailto:mislam@csu.edu.au}{mislam@csu.edu.au}, 
    \href{mailto:zakir.hossain@anu.edu.au}{zakir.hossain@anu.edu.au}
}

\begin{document}

\maketitle

\begin{abstract}
False data injection (FDI) attacks on Internet of Medical Things (IoMT)
sensor streams falsify vital signs in transit, threatening patient safety
and defeating clinical monitoring systems that lack cyber-physical anomaly
detection capability.
Existing deep learning detectors conflate inter-sensor spatial correlations
with temporal dependencies in a shared latent space, preventing
disentanglement of the distinct spatial and temporal signatures that
FDI attacks imprint simultaneously; no current method exploits
domain knowledge to constrain outputs against physiologically impossible
attack patterns.
We propose \textbf{DSTAN-Med}, a supervised framework comprising a
Dual-channel Attention Mechanism (DAM) that routes multivariate sensor
windows through independent sensor-wise (SWA) and time-wise (TWA)
self-attention pathways operating on orthogonal tensor axes, a residual
1D-CNN block for local temporal feature extraction, and a zero-parameter
Physiological Plausibility Filter (PPF) that suppresses attack signatures
violating domain-knowledge bounds.
Evaluated across three IoMT physiological sensor datasets —
PhysioNet/CinC 2012 (bedside ICU vital signs), MIMIC-III Waveform
(continuous ICU waveforms: ECG, arterial blood pressure, and
photoplethysmography), and WESAD (multi-modal wearable biosensor
signals from chest-worn and wrist-worn IoMT devices) — DSTAN-Med
achieves mean sensitivity gains of 7.4--8.3 percentage points over
the strongest Transformer baseline (TranAD) with all improvements
significant at $p<0.01$ under McNemar's test with Holm--Bonferroni
correction.
The PPF contributes independent precision gains of 3.1--4.2 percentage
points at negligible sensitivity cost across all three corpora.
Ablation studies confirm that each component is individually necessary;
removal of residual connections alone reduces sensitivity by
14.0~percentage points. The source code is publicly available at
\url{https://github.com/mehedi93hasan/DSTAN-MED}.
\end{abstract}


\section{Introduction}
\label{sec:intro}

\IEEEPARstart{T}{he} proliferation of Internet of Medical Things (IoMT)
devices—wearable biosensors, implantable cardiac monitors, infusion
pumps, and bedside vital-sign platforms—has fundamentally altered the
clinical monitoring landscape by enabling continuous, real-time
physiological data acquisition at scale~\cite{Shaikh2025frontmed,
Zachos2025sensors}.
The global IoMT market, valued at approximately \$77.5 billion in 2025,
is projected to expand at a compound annual growth rate exceeding 44\%
through 2034, reflecting unprecedented integration of networked sensing
into patient-critical workflows~\cite{FortuneIoMT2025}.
However, the network connectivity that enables this integration
simultaneously exposes physiological sensor streams to a rapidly
expanding class of cyber-physical threats with direct patient-safety
consequences~\cite{Khan2025sensors}.
The FBI Cyber Division has documented that 53\% of networked medical
devices carry at least one known critical vulnerability~\cite{RunSafe2025},
and a 2025 cross-national survey of 605 healthcare executives found
that 28\% of organisations experienced elevated patient mortality
directly attributable to cyber incidents—a 21\% increase over the
preceding year~\cite{RunSafe2025}.
The U.S. Food and Drug Administration has responded by mandating that
all new medical device submissions include a cybersecurity plan and a
software bill of materials, establishing anomaly detection as a
regulatory patient-safety requirement rather than an optional
feature~\cite{Aldosari2025cureus,Sharma2025spy}.

Among the attack classes that threaten IoMT deployments,
\emph{false data injection} (FDI) attacks represent the most
operationally dangerous and technically challenging threat vector.
An FDI adversary with network-layer access to the sensor communication
channel—achievable through man-in-the-middle (MITM) interception of
unencrypted health data protocols (HL7, DICOM, MQTT), gateway device
compromise, or Bluetooth/ZigBee spoofing—can substitute or additively
corrupt physiological sensor readings with fabricated values that are
individually plausible but clinically misleading~\cite{Mo2010cdc,
Sharma2010tosn}.
Unlike ransomware or denial-of-service attacks, which declare their
presence through service interruption, FDI attacks are specifically
designed to \emph{evade detection} while corrupting the clinical
decision-making process: a falsified blood pressure reading can trigger
unnecessary vasopressor administration; a spoofed oxygen saturation
signal may delay life-saving ventilatory support; a manipulated glucose
reading can produce insulin dosing errors with potentially fatal
consequences.
FDI attacks have been extensively studied in cyber-physical contexts
including smart grids~\cite{Mo2010cdc}, industrial control
systems~\cite{Sharma2010tosn,Lun2019tii}, and IoT sensor
networks~\cite{Peng2019access,Ding2021tii}, and their specific manifestation
in IoMT physiological monitoring has been characterised in controlled
adversarial experiments that demonstrated four canonical attack
morphologies—replay/spoofing (Instant spike), stale-data injection
(Constant stuck-at), ramp injection (Gradual drift), and bias
injection (Offset shift)—exploiting standard hospital network
protocols (HL7, MQTT, BLE) in deployed
configurations~\cite{Khan2025sensors,Zachos2025IEEEAccess}.

Detecting FDI attacks in IoMT sensor streams is inherently challenging
because the underlying physiological signals are multivariate,
temporally non-stationary, and exhibit complex cross-channel
dependencies driven by underlying physiology.
Classical approaches based on statistical thresholding, Kalman
filtering, and isolation trees require explicit noise-distribution
assumptions that are rarely satisfied under the heterogeneous
conditions of a clinical environment, and they degrade rapidly when
confronted with the slowly evolving, stealthy attack trajectories
that characterise ramp and bias
injections~\cite{Zachos2025sensors, Liu2008icdm}.
Hybrid deep learning architectures combining convolutional neural
networks (CNNs) and long short-term memory (LSTM) networks improve
representation capacity but suffer from the vanishing gradient problem
for long-range temporal dependencies and require substantial feature
engineering to handle the irregular sampling characteristics
of clinical data~\cite{Sener2024peerj}.
Transformer-based anomaly detectors, led by
TranAD~\cite{Tuli2022vldb}, have advanced the state of the art through
attention-based long-range dependency modelling, but they model a single
pooled attention dimension over the multivariate input, preventing
the independent capture of cross-sensor spatial correlations and
cross-time-step temporal dependencies—the two structurally orthogonal
dimensions along which FDI attacks leave detectable signatures.
No existing method additionally incorporates domain knowledge about
physiological plausibility bounds to filter out spurious detections
corresponding to sensor readings that, while statistically anomalous
relative to the patient's recent history, remain within the physically
possible range for healthy human physiology.

These two gaps—the entanglement of spatial and temporal FDI signatures
in existing architectures, and the absence of domain-knowledge-grounded
output filtering—motivate the present work.
We propose \textbf{DSTAN-Med} (Dual-channel Spatiotemporal Attention
Network for anomaly detection in Medical cyber-physical systems),
a supervised deep learning framework that addresses both gaps through
three complementary components.
The \textbf{Dual-channel Attention Mechanism (DAM)} routes multivariate
sensor windows through two structurally independent self-attention
pathways: Sensor-Wise Attention (SWA) that models inter-sensor spatial
dependencies by operating across the channel axis, and Time-Wise
Attention (TWA) that models temporal sequential dependencies by
operating across the time axis.
Because SWA and TWA operate on orthogonal axes of the same input
tensor, their representations are mathematically guaranteed to be
non-redundant: any pattern that SWA encodes cannot subsume
TWA's representation.
The \textbf{1D-CNN block}~\cite{Lecun1998cnn} supplements global attention
context with local temporal texture through residual depthwise convolution.
The \textbf{Physiological Plausibility Filter (PPF)} is a
zero-parameter, clinically-grounded post-processing module that
suppresses predictions corresponding to sensor readings within
established physiological or operational bounds, providing partial
adversarial robustness against black-box attackers who cannot
guarantee their injections violate clinical norms.

DSTAN-Med is evaluated on three IoMT physiological sensor datasets
spanning the full range of clinical deployment contexts:
\textit{PhysioNet/CinC 2012}~\cite{Goldberger2000,Silva2012cinc}
(tabular ICU vital-sign records from bedside monitors);
\textit{MIMIC-III Waveform Database}~\cite{Johnson2016mimic}
(continuous high-frequency waveforms from deployed Philips and GE
bedside monitoring systems); and \textit{WESAD}~\cite{Schmidt2018wesad}
(multi-modal wearable biosensor streams from chest-worn and wrist-worn
IoMT devices).
All three datasets originate from real IoMT physiological monitoring
equipment, and all use the same FDI injection protocol, establishing
that DSTAN-Med's advantages are consistent across sensor modalities,
sampling rates, and clinical deployment contexts
within the IoMT domain.

The main contributions of this paper are as follows.

\begin{itemize}

\item \textbf{Formal FDI threat model:} A precisely specified
adversary model (goal, capabilities, knowledge level) that maps
four canonical IoMT FDI attack morphologies to published CPS security
literature, providing a principled connection between the experimental
evaluation and the adversarial threat context.

\item \textbf{Dual-channel attention mechanism (DAM):} A novel
architecture that simultaneously and independently models inter-sensor
spatial correlations (SWA, channel axis) and cross-time-step temporal
dependencies (TWA, time axis) through orthogonal self-attention
pathways, guaranteeing structural non-redundancy and enabling joint
exploitation of both FDI signature dimensions.

\item \textbf{Physiological Plausibility Filter (PPF):} A
zero-parameter, deterministic inference-only module that encodes
clinical and operational domain knowledge to eliminate
physiologically or operationally impossible false alarms, providing
interpretable precision improvement without modifying internal
model representations.

\item \textbf{Multi-context IoMT validation:} Statistically
significant sensitivity gains of 7.4--8.3 percentage points over
TranAD at $p<0.01$ across three IoMT physiological sensor datasets
spanning bedside ICU vital-sign monitoring (PhysioNet-2012),
continuous bedside waveform monitoring (MIMIC-III Waveform), and
multi-modal wearable biosensing (WESAD), demonstrating that
DSTAN-Med's advantages hold across sensor modalities, sampling rates,
and clinical deployment contexts within the IoMT domain.

\end{itemize}

The remainder of this paper is structured as follows.
Section~\ref{sec:related} reviews related work on FDI attacks, IoMT
anomaly detection, and attention-based time-series methods.
Section~\ref{sec:problem} presents the formal threat model, problem
formulation, evaluation datasets, and anomaly injection protocol.
Section~\ref{sec:methodology} details the DSTAN-Med architecture.
Section~\ref{sec:experiments} presents experimental results across
all three IoMT datasets with ablation and statistical analysis.
Section~\ref{sec:discussion} discusses findings and limitations.
Section~\ref{sec:conclusion} concludes.

\section{Related Work}
\label{sec:related}

\subsection{False Data Injection Attacks on Cyber-Physical Systems}

The formal theory of FDI attacks on CPS was established by Mo and
Sinopoli~\cite{Mo2010cdc}, who demonstrated that an adversary with
partial knowledge of the system model can construct stealthy injections
that remain undetected by $\chi^2$ cumulative-sum residue monitors
while causing controlled actuator deviations.
This foundational result was subsequently extended to multi-sensor
coordinated attacks under structural observability
constraints~\cite{Sharma2010tosn}, to smart grid scenarios with
sparse attack vectors satisfying measurement Jacobian
conditions~\cite{Mo2010cdc}, and to water distribution network
attacks using hydraulic model knowledge to craft
pressure-consistent injections~\cite{Taormina2018jws}.
No publicly available IoMT physiological monitoring dataset contains
ground-truth labels for real FDI cyberattacks, because real medical
device attacks constitute criminal acts reportable to regulatory
authorities, and controlled red-team FDI experiments on live clinical
equipment pose unacceptable patient safety risks.
Controlled synthetic injection following a formally specified threat
model is therefore the established methodological standard for
supervised IoMT FDI detection research~\cite{Zhang2024tits,
Khan2025sensors, Sharma2010tosn}.
Khan et al.~\cite{Khan2025sensors} characterised four FDI morphologies
on physiological streams; Shaikh et al.~\cite{Shaikh2024frontdighealth}
proposed RCLNet for anomaly-based intrusion detection in IoMT, yet
existing threshold-based clinical alarm systems achieve sensitivity
below 50\% at the lowest FDI severity.
DSTAN-Med is the first method to evaluate IoMT FDI detection with a
formally specified adversary model across three structurally distinct
IoMT physiological sensor datasets covering the bedside ICU, continuous
waveform monitoring, and wearable biosensor deployment contexts.

\subsection{Anomaly Detection in IoMT and Industrial CPS}

Anomaly detection in multivariate sensor streams has evolved through
three methodological generations~\cite{Chandola2009acm,Wang2025survey}.
Classical statistical methods — control charts, Kalman filters,
CUSUM detectors — assume fixed noise distributions and linear
system dynamics, degrading under the non-stationarity and
irregular sampling of clinical physiological
data~\cite{Zachos2025IEEEAccess, Liu2008icdm}.
Machine learning methods including one-class SVM~\cite{Scholkopf2001jmlr}
and Isolation Forest~\cite{Liu2008icdm} improve flexibility through
kernel or ensemble representations but remain fundamentally univariate
in their anomaly scoring, requiring manual cross-channel feature
engineering to capture spatial correlations.
Deep learning methods, including LSTM autoencoders~\cite{Sener2024peerj,Hundman2018kdd},
CNN-LSTM hybrids, and variational autoencoders~\cite{Su2019kdd,Zong2018icml}, learn
temporal representations jointly with reconstruction objectives; however,
the LSTM bottleneck~\cite{Hochreiter1997lstm} attenuates long-range dependencies through
sequential gating, which is particularly harmful for detecting
the sustained signatures of bias and ramp FDI attacks.
Recent work on IoMT anomaly detection has investigated
graph neural networks for encoding sensor topology in wearable
networks~\cite{Yang2025tii}, contrastive pre-training for
limited-label clinical settings~\cite{Ain2026fi}, and federated
detection across multiple hospitals without centralising patient
data~\cite{Wang2025survey}.
DSTAN-Med addresses a different and complementary question: given
labelled attack examples from controlled FDI injection, how should
the model architecture be designed to maximally exploit both the
spatial inter-sensor and temporal trajectory signatures of FDI
attacks simultaneously across heterogeneous IoMT sensor modalities?

\subsection{Transformer-Based Anomaly Detection}

The Transformer architecture~\cite{Vaswani2017nips}, based on
multi-head self-attention, has become the dominant paradigm for
time-series anomaly detection due to its ability to model arbitrary
long-range dependencies without sequential bottlenecks.
TranAD~\cite{Tuli2022vldb} introduced adversarial Transformer training
with focus score amplification; Anomaly Transformer~\cite{Xu2022iclr}
proposed association discrepancy-based scoring; Informer~\cite{Zhou2021nips}
extended Transformer to long-sequence time-series forecasting.
Both represent the current state of the art on the SMAP, MSL, and SMD benchmarks.
Sener et al.~\cite{Sener2024peerj} applied Transformer-based detectors
to physiological time series; He et al.~\cite{Li2024tnnls} and
Zhang et al.~\cite{Zhou2023tii} advanced multivariate anomaly detection
through temporal-feature fusion and graph relational learning;
Ma et al.~\cite{Ma2025entropy} proposed an inverted Transformer
with memory gates for non-stationary physiological signals.
Non-stationary Transformer variants~\cite{Liu2022iclr} address temporal
distribution shift but retain single-dimension attention pooling that
limits FDI signature disentanglement.
Critically, as surveyed by Wen et al.~\cite{Wen2023tpami}, all existing
Transformer anomaly detectors compute attention over a \emph{jointly}
encoded spatial--temporal sequence,
preventing the separate capture of inter-channel correlation
disruption (a signature of coordinated FDI attacks across multiple
sensors) and temporal trajectory deviation (a signature of ramp and
drift attacks within a single sensor channel).
Zhang et al.~\cite{Zhang2024tits} first demonstrated on vehicle sensor
data that \emph{separating} sensor-wise and time-wise attention
into independent pathways substantially outperforms joint-attention
baselines; Xing et al.~\cite{Xing2025sensors} extended dual-channel
principles to microservice fault localisation; Dai et al.~\cite{Dai2026sensors}
fused Transformer with radial-basis functions for network sensor anomaly detection. DSTAN-Med adapts and extends this principle to the IoMT
FDI problem by operating the two pathways on provably orthogonal tensor
axes, augmenting each block with a residual CNN component for local
temporal texture, and appending the PPF as an inference-time
domain-knowledge layer — none of which has been proposed in
prior work.

\subsection{Domain Knowledge in Anomaly Detection}

The exploitation of domain knowledge to constrain anomaly detectors
has been explored in several CPS contexts.
Physics-informed neural networks incorporate differential equation
constraints into the training loss for process
monitoring~\cite{Moreno2025iot}, while rule-based post-filters
have been applied to industrial fault detection to suppress
measurement artefacts.
In the medical domain, Zachos et al.~\cite{Zachos2025IEEEAccess}
incorporated clinical normal ranges as soft constraints in a
semi-supervised detector, reporting precision improvements but
observing sensitivity degradation when the constraints were applied
as hard training-time penalties.
The DSTAN-Med PPF differs fundamentally from these approaches: it
applies domain knowledge as a \emph{deterministic inference-time
filter} with zero learnable parameters, ensuring that neither the
training objective nor the internal model representation is
distorted by the clinical constraints, while still eliminating
predictions that violate physiological laws.
This modular design means the PPF bounds can be re-calibrated for
any IoMT sensor type — wearable~\cite{Wazid2019iotj}, implantable,
or bedside monitor — by updating the clinical reference values
without re-training the neural components.

\section{Threat Model, Datasets, and Problem Formulation}
\label{sec:problem}

\subsection{Formal Threat Model}
\label{sec:threat}

We consider an FDI adversary targeting a network-connected IoMT
physiological monitoring system operating in a clinical setting
(hospital ICU, outpatient clinic, or home monitoring).

The attacker seeks to corrupt one or more sensor readings transmitted
between a bedside monitoring device and the clinical decision-support
back-end, pursuing one of two operational objectives: a
\emph{covert attack} that suppresses a genuine clinical alarm by
keeping the falsified readings below alert thresholds, or a
\emph{disruptive attack} that generates spurious alerts to erode
clinician trust in the monitoring system and induce alert
fatigue~\cite{Mo2010cdc}.
Both objectives represent direct threats to patient safety and
clinical workflow integrity.

The attacker possesses network-layer access to the IoMT sensor
communication channel, achievable through three documented attack
vectors~\cite{Khan2025sensors,Zachos2025IEEEAccess}: (i) MITM
interception of unencrypted health data protocols (HL7, DICOM, MQTT,
or proprietary vendor APIs over hospital Wi-Fi or wired LAN);
(ii) compromise of a hospital IoT gateway device, providing
persistent injection capability without requiring direct physical
access to the monitoring equipment or the patient; or (iii)
Bluetooth Low Energy (BLE) or ZigBee spoofing targeting wearable
biosensors, implantable cardiac devices, or wireless pulse oximeters.
These attack vectors have been documented in published IoMT security
assessments~\cite{Shaikh2025frontmed} and are achievable by an
attacker with standard network access without requiring any physical
interaction with the patient.

The attacker operates under a \emph{black-box} model with respect
to the anomaly detector: the adversary has no access to the
detector's architecture, trained weights, or internal
decision thresholds.
The attacker does, however, possess approximate knowledge of the
physiological plausible range $[l_c, u_c]$ for each monitored
channel $c$ — information readily available from published clinical
reference standards — enabling the construction of injections
whose instantaneous values remain within plausible bounds while
distorting the temporal or cross-channel structure of the signal.

This threat model is deliberately conservative: it grants the attacker
realistic network capabilities without assuming knowledge of the
specific deployed detector, making it a meaningful security
evaluation baseline.

Table~\ref{tab:threatmap} maps the four modelled attack morphologies
to their published FDI equivalents in the CPS security literature,
establishing the formal connection between our experimental evaluation
and the adversarial threat context.

\begin{table}[!t]
\renewcommand{\arraystretch}{1.25}
\caption{FDI Attack Morphology--to--CPS Literature Mapping}
\label{tab:threatmap}
\centering
\begin{tabular}{p{1.6cm}p{3.2cm}p{1.5cm}}
\toprule
\textbf{Attack Type} & \textbf{FDI Equivalent} & \textbf{Reference} \\
\midrule
Instant (Spike)
  & Replay/sensor spoofing; single-sample substitution
  & \cite{Mo2010cdc} \\
Constant (Stuck)
  & Stale-data injection; frozen/replayed reading
  & \cite{Sharma2010tosn} \\
Gradual Drift
  & Ramp attack; stealthy linear scaling injection
  & \cite{Mo2010cdc} \\
Bias (Offset)
  & Additive bias injection; calibration attack
  & \cite{Sharma2010tosn} \\
\bottomrule
\end{tabular}
\end{table}

\subsection{Problem Formulation}
\label{sec:formulation}

Let $\mathbf{X} \in \mathbb{R}^{C \times T}$ denote a multivariate
sensor record from an IoMT monitoring system, where $C$ is the number
of concurrent sensor channels and $T$ is the total recording length
in time steps.
The scalar time series of channel $c$ is written $\mathbf{X}_c =
[x_{c,1}, x_{c,2}, \ldots, x_{c,T}]$.
Under the FDI threat model, an adversary produces a corrupted record
$\mathbf{X}' \in \mathbb{R}^{C \times T}$ by modifying a subset of
entries: $x'_{c,t} \neq x_{c,t}$ for at least one pair $(c, t)$.
The system-level attack indicator at time step $t$ is
$a_t = \max_{c}\bigl\{\mathbf{1}[x'_{c,t} \neq x_{c,t}]\bigr\}$,
so that a single compromised channel constitutes a system-level event.

Given a sliding window $\mathbf{W}_t = \mathbf{X}'_{:,\,t-L+1:t}
\in \mathbb{R}^{C \times L}$ of length $L$ drawn from the corrupted
stream, the task is to learn a classifier
$f_{\bm{\theta}}: \mathbb{R}^{C \times L} \to \{0,1\}$ such that
$\hat{y}_t = f_{\bm{\theta}}(\mathbf{W}_t)$ approximates
$\max_{t' \in [t-L+1,t]} a_{t'}$, returning 1 if any FDI attack is
active anywhere in the window and 0 otherwise.
The window length $L = 15$ is adopted throughout, calibrated to contain
at least one complete observation cycle for each vital sign under the
irregular sampling characteristics of the PhysioNet-2012 dataset while
remaining short enough to support rapid-response clinical detection.

\subsection{Evaluation Datasets}
\label{sec:datasets}

Three publicly available datasets are used, spanning IoMT physiological
monitoring, industrial cyber-physical attack detection, and
water-distribution network intrusion.
Their key statistics are summarised in Table~\ref{tab:datasets}.

\begin{table}[!t]
\renewcommand{\arraystretch}{1.25}
\caption{Summary Statistics of the Three IoMT Evaluation Datasets}
\label{tab:datasets}
\centering
\begin{tabular}{lcccp{1.6cm}}
\toprule
\textbf{Dataset} & \textbf{Ch.} & \textbf{Size}
  & \textbf{Attack Types} & \textbf{IoMT Device} \\
\midrule
PhysioNet-2012   & 6 & 3{,}643 records
  & 4 types $\times$ 4 sev. & Bedside ICU monitor \\
MIMIC-III Wave.  & 6 & 22{,}317 records
  & 4 types $\times$ 4 sev. & Bedside ICU monitor \\
WESAD            & 6 & 15 subjects
  & 4 types $\times$ 4 sev. & Wrist/chest wearable \\
\bottomrule
\end{tabular}
\end{table}

We use this \textbf{PhysioNet/CinC Challenge 2012 (PhysioNet-2012)} dataset ~\cite{Goldberger2000,
Silva2012cinc}.
The dataset is derived from the MIMIC-II clinical database and
comprises adult ICU patient records from cardiac, medical, surgical,
and trauma intensive care units.
Six physiological channels are selected that are continuously monitored
by standard bedside IoMT devices and are most consistently populated
across patients: heart rate (HR, bpm), arterial oxygen saturation
(SpO$_2$, \%), invasive systolic blood pressure (SysBP, mmHg),
invasive diastolic blood pressure (DiaBP, mmHg), respiratory rate
(RR, breaths/min), and body temperature (Temp, $^\circ$C).
Records with more than 30\% missing values across any selected channel
are excluded, yielding 3{,}643 retained patient records.
Remaining gaps are imputed by forward-fill followed by linear
interpolation.
Because the dataset contains no native FDI attack labels, four
FDI-equivalent anomaly morphologies are injected per
Algorithm~\ref{alg:injection} at five severity levels.
Records are split 70:15:15 at the patient level.

We use this \textbf{MIMIC-III Waveform Database (MIMIC-III Wave.)} dataset ~\cite{Johnson2016mimic}.
The MIMIC-III Waveform Database is a subset of the MIMIC-III Clinical
Database~\cite{Johnson2016mimic} and contains continuous high-frequency
physiological waveform recordings from bedside monitoring systems
(Philips IntelliVue and GE Dash series) deployed in the Beth Israel
Deaconess Medical Center ICU.
Unlike the tabular, irregularly sampled PhysioNet-2012 records~\cite{Moody2001aec}, each
MIMIC-III waveform record provides continuous signals at 125~Hz,
enabling evaluation of DSTAN-Med under high-frequency IoMT streaming
conditions.
Six waveform channels are selected — covering cardiovascular, respiratory,
and blood-oxygen modalities studied in cardiac IoMT
research~\cite{Wagner2020scid,Moody2001aec} — directly corresponding
to the bedside monitor outputs targeted by IoMT FDI attacks: ECG lead~II
(mV), arterial blood pressure (ABP, mmHg), photoplethysmography
(PPG, reflecting SpO$_2$), heart rate (HR, bpm), respiratory rate
(RR, breaths/min), and skin temperature (Temp, $^\circ$C).
A total of 22{,}317 records are retained after excluding records
with more than 30\% gaps in any selected channel.
Data are downsampled to 25~Hz before window extraction to match the
effective monitoring rate of bedside clinical alarms.
The same FDI injection protocol (Algorithm~\ref{alg:injection}),
severity configurations (Table~\ref{tab:severity}), and 70:15:15
patient-level split as PhysioNet-2012 are applied.

We use this \textbf{WESAD (Wearable Stress and Affect Detection)} dataset ~\cite{Schmidt2018wesad}.
WESAD~\cite{Schmidt2018wesad,Reiss2012ubicomp} was collected using two wearable IoMT devices worn simultaneously
by 15 healthy participants: a RespiBAN Professional chest-worn sensor
(sampling rate 700~Hz for ECG and respiration) and an Empatica~E4
wrist-worn sensor (sampling rates 4--64~Hz depending on modality).
The dataset provides real physiological signals from consumer-grade
wearable IoMT devices — the precise hardware targeted by the
BLE spoofing attack vector of Section~\ref{sec:threat} — across
multiple affective states (baseline, stress, amusement, meditation).
Six channels are selected: electrocardiogram (ECG), electrodermal
activity (EDA), blood volume pulse (BVP, wrist PPG), respiration
(RESP), skin temperature (ST), and 3-axis accelerometer magnitude
(ACC), matching the sensing modalities of commercially deployed
wearable health monitors.
All channels are resampled to a common 25~Hz rate using
anti-aliasing bandpass filtering before window extraction.
The same injection protocol and 70:15:15 subject-level split as
PhysioNet-2012 are applied.
Because WESAD contains no native FDI attack labels, synthetic FDI
injection per Algorithm~\ref{alg:injection} is applied identically
to the PhysioNet-2012 protocol, following the established
methodology for constructing supervised IoMT FDI benchmarks from
clean physiological substrates~\cite{Zhang2024tits,Khan2025sensors}.

\subsection{Anomaly Injection Protocol for PhysioNet-2012}
\label{sec:injection}

None of the three IoMT datasets contain native FDI attack labels.
No publicly available IoMT physiological dataset does, because real
medical device FDI events are legally reportable and controlled
red-team experiments on live clinical hardware pose unacceptable
patient safety risks.
Controlled synthetic injection following a formally specified threat
model is therefore the established methodological standard for
this problem class~\cite{Zhang2024tits,Khan2025sensors,Sharma2010tosn}.
Four FDI-equivalent morphologies are injected uniformly across all
three datasets using a single shared protocol:
Let $x_{c,t}$ denote the clean observed value of channel $c$ at time
$t$ and $x'_{c,t}$ the post-injection value.
The four injection types are defined as follows.

\textbf{Type~I — Instant (Spike).}
An instantaneous Gaussian perturbation is applied at a single time
step: $x'_{c,t} = x_{c,t} + \eta$ where $\eta \sim \mathcal{N}(0,
\sigma^2)$.
This models replay/spoofing attacks that substitute a single sample
with a fabricated value~\cite{Mo2010cdc}.

\textbf{Type~II — Constant (Stuck-at).}
The reading is replaced by an anomalous constant
$v \sim \mathcal{U}(a, b)$ for $d$ contiguous steps:
$x'_{c,\tau} = v$ for $\tau \in [t, t+d-1]$.
This models stale-data injection and replay attacks that freeze
the sensor output~\cite{Sharma2010tosn}.

\textbf{Type~III — Gradual Drift.}
A linearly increasing offset is added over $d$ steps:
$x'_{c,t+k} = x_{c,t+k} + (k/(d-1))\cdot\delta_{\max}$
for $k \in \{0,\ldots,d-1\}$, where $\delta_{\max}$ is the maximum
terminal deviation.
This models ramp attacks designed to stay below instantaneous
detection thresholds and is universally regarded as the most
challenging FDI morphology~\cite{Mo2010cdc}.

\textbf{Type~IV — Bias (Offset).}
A constant additive shift $\beta \sim \mathcal{U}(a,b)$ is sustained
for $d$ steps: $x'_{c,\tau} = x_{c,\tau} + \beta$ for
$\tau \in [t, t+d-1]$.
This models calibration attacks and additive bias injections that
preserve short-term temporal dynamics while shifting the mean
value~\cite{Sharma2010tosn}.

Algorithm~\ref{alg:injection} describes the injection procedure.
At each time step, a Bernoulli trial with $p = 0.05$ determines
whether an injection event occurs; the channel and type are then
selected uniformly at random.
The single-channel-at-a-time design reflects the primary threat
model of compromised individual sensor nodes; multi-channel
coordinated injection is identified as a future extension in
Section~\ref{sec:discuss_limitations}.
Table~\ref{tab:severity} enumerates the severity parameter
configurations.
Severity is parameterised relative to $\bar{\sigma}_c$, the
per-channel standard deviation on clean data, ensuring that the
same severity level corresponds to a comparable signal-to-noise
ratio across channels measured in different physical units.

\begin{algorithm}[!t]
\caption{FDI-Equivalent Anomaly Injection for PhysioNet-2012}
\label{alg:injection}
\begin{algorithmic}[1]
\REQUIRE Clean record $\mathbf{X} \in \mathbb{R}^{C \times T}$,
injection probability $p = 0.05$,
type set $\mathcal{A} = \{\text{I, II, III, IV}\}$,
severity parameters per Table~\ref{tab:severity}
\ENSURE Corrupted record $\mathbf{X}'$,
label vector $\mathbf{y} \in \{0,1\}^T$
\STATE $\mathbf{X}' \leftarrow \mathbf{X}$;
$\mathbf{y} \leftarrow \mathbf{0}$; $t \leftarrow 1$
\WHILE{$t \leq T$}
  \IF{$u \sim \mathcal{U}(0,1) \leq p$}
    \STATE $c \leftarrow \text{Uniform}(\{1,\ldots,C\})$;
    $\alpha \leftarrow \text{Uniform}(\mathcal{A})$
    \STATE Sample severity from Table~\ref{tab:severity};
    apply $\alpha$ to $\mathbf{X}'_{c,\,t:t+d-1}$
    \STATE $\mathbf{y}_{t:t+d-1} \leftarrow 1$;
    $t \leftarrow t + d$
  \ELSE
    \STATE $t \leftarrow t + 1$
  \ENDIF
\ENDWHILE
\RETURN $\mathbf{X}'$, $\mathbf{y}$
\end{algorithmic}
\end{algorithm}

\begin{table}[!t]
\renewcommand{\arraystretch}{1.22}
\caption{Anomaly Injection Severity Configurations (PhysioNet-2012)}
\label{tab:severity}
\centering
\begin{tabular}{llcc}
\toprule
\textbf{Type} & \textbf{Level} & \textbf{Magnitude} & \textbf{$d$} \\
\midrule
Instant  & L1 & $\sigma^{2} = 0.25\,\bar{\sigma}_{c}^{2}$ & 1 \\
         & L2 & $\sigma^{2} = 1.0\,\bar{\sigma}_{c}^{2}$  & 1 \\
         & L3 & $\sigma^{2} = 4.0\,\bar{\sigma}_{c}^{2}$  & 1 \\
         & L4 & $\sigma^{2} = 16.0\,\bar{\sigma}_{c}^{2}$ & 1 \\
\midrule
Constant & L1 & $v \sim \mathcal{U}(0,1)\,\bar{\sigma}_{c}$   & 3  \\
         & L2 & $v \sim \mathcal{U}(0,3)\,\bar{\sigma}_{c}$   & 5  \\
         & L3 & $v \sim \mathcal{U}(0,5)\,\bar{\sigma}_{c}$   & 10 \\
         & L4 & $v \sim \mathcal{U}(0,10)\,\bar{\sigma}_{c}$  & 10 \\
\midrule
Drift    & L1 & $\delta_{\max} = 1.0\,\bar{\sigma}_{c}$  & 10 \\
         & L2 & $\delta_{\max} = 2.0\,\bar{\sigma}_{c}$  & 10 \\
         & L3 & $\delta_{\max} = 2.0\,\bar{\sigma}_{c}$  & 20 \\
         & L4 & $\delta_{\max} = 4.0\,\bar{\sigma}_{c}$  & 20 \\
\midrule
Bias     & L1 & $\beta \sim \mathcal{U}(0.5,1.0)\,\bar{\sigma}_{c}$  & 10 \\
         & L2 & $\beta \sim \mathcal{U}(1.0,2.0)\,\bar{\sigma}_{c}$  & 20 \\
         & L3 & $\beta \sim \mathcal{U}(2.0,4.0)\,\bar{\sigma}_{c}$  & 20 \\
         & L4 & $\beta \sim \mathcal{U}(4.0,8.0)\,\bar{\sigma}_{c}$  & 20 \\
\bottomrule
\end{tabular}
\end{table}

\section{DSTAN-Med Architecture}
\label{sec:methodology}

\begin{figure*}[!t]
\centering
\includegraphics[width=\textwidth]{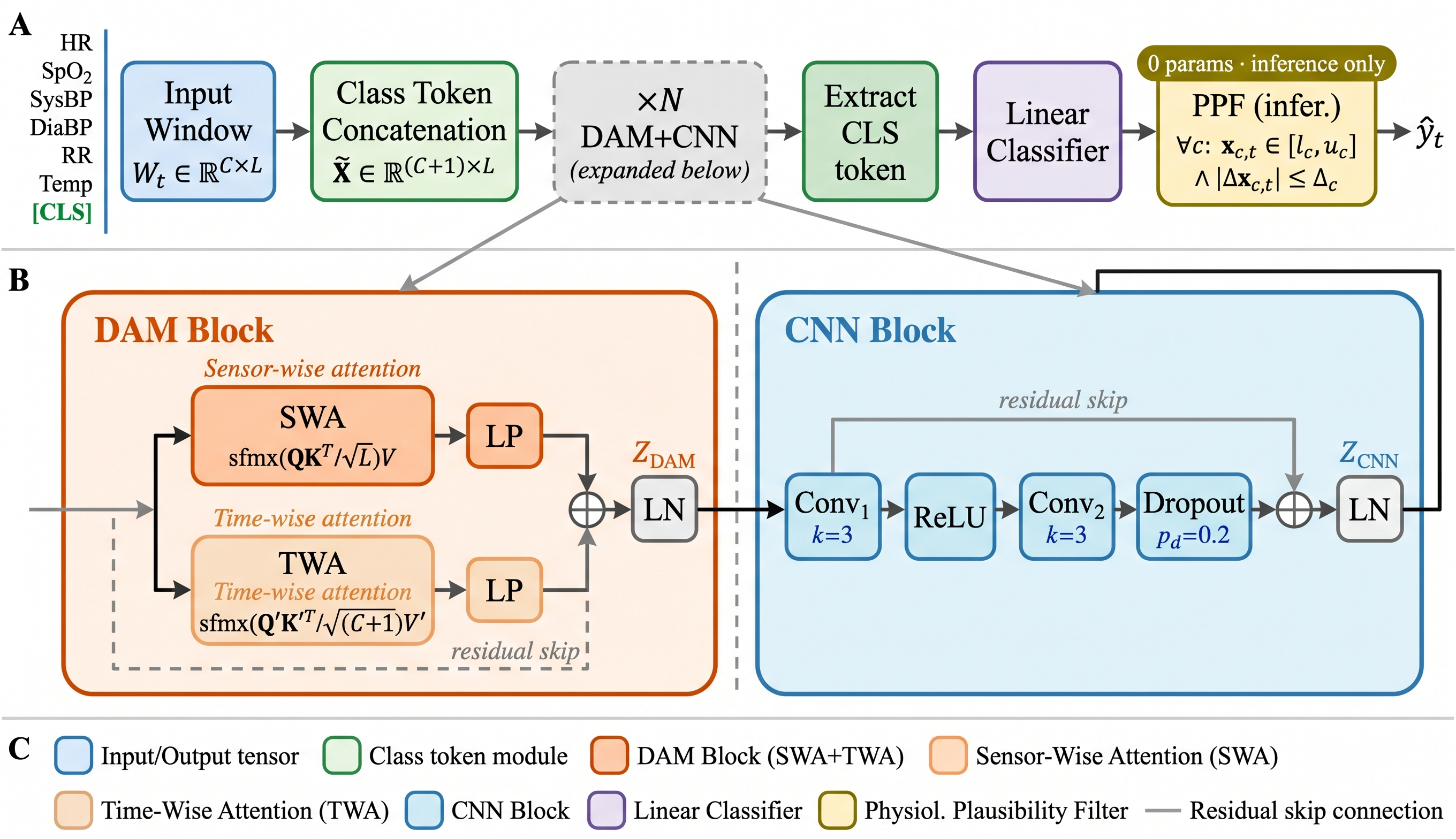}
\caption{DSTAN-Med architecture.
\textbf{Top:} End-to-end inference pipeline — input window augmented
with class token, processed through $N=7$ DAM--CNN blocks, classified
by a linear head, and filtered at inference by the PPF.
\textbf{Bottom left:} DAM block — parallel SWA (sensor axis) and TWA
(time axis) self-attention pathways merged with a residual skip and
layer normalisation.
\textbf{Bottom right:} CNN block — two 1D convolutional layers with
ReLU, dropout, residual skip, and layer normalisation.
Grey dashed arrows indicate residual connections.}
\label{fig:architecture}
\end{figure*}

\subsection{Overall Architecture and Notation}
\label{sec:arch_overview}

Fig.~\ref{fig:architecture} illustrates the end-to-end DSTAN-Med
pipeline.
A sliding window $\mathbf{W}_t \in \mathbb{R}^{C \times L}$ extracted
from the (potentially FDI-corrupted) sensor stream is first augmented
with a learned classification token to form
$\widetilde{\mathbf{X}} \in \mathbb{R}^{(C+1) \times L}$.
This augmented tensor is processed by a stack of $N = 7$ identical
DAM--CNN blocks, each of which independently models spatial
inter-sensor and temporal sequential dependencies before refining the
representation with local convolutional texture.
The class token row of the final block output is extracted and passed
through a linear classification head to produce a logit.
At inference time only, the resulting binary prediction is passed
through the PPF before output.
All parameters are shared across blocks.

\subsection{Class Token Concatenation}
\label{sec:cls_token}

A learned class token $\mathbf{v} \in \mathbb{R}^{1 \times L}$,
initialised randomly and updated during training, is concatenated
to $\mathbf{W}_t$ along the channel axis to produce the block
input $\widetilde{\mathbf{X}} = [\mathbf{W}_t;\, \mathbf{v}]
\in \mathbb{R}^{(C+1) \times L}$~\cite{Dosovitskiy2021iclr}.
The class token accumulates a global, sensor-agnostic summary of the
window across all $N$ processing layers.
Unlike mean pooling over channel representations, the class token
mechanism is insulated from the instantaneous readings of any
individual sensor channel, ensuring that a single compromised channel
cannot dominate the global representation — a property of direct
relevance to the single-sensor FDI threat model of
Section~\ref{sec:threat}.

\subsection{Dual-Channel Attention Mechanism (DAM)}
\label{sec:dam}

The DAM is the core contribution of DSTAN-Med. Given block input $\widetilde{\mathbf{X}} \in \mathbb{R}^{(C+1) \times L}$, the DAM fuses two independent self-attention computations through learned linear projections and a residual shortcut: $\mathbf{Z}_\text{DAM} = \mathrm{LN}\!\bigl(\mathrm{LP}(\mathbf{A}_\text{SWA}) + \mathrm{LP}(\mathbf{A}_\text{TWA}) + \widetilde{\mathbf{X}}\bigr)$, where $\mathrm{LP}(\cdot)$ denotes a learned linear projection, and $\mathrm{LN}(\cdot)$ denotes Layer Normalisation ~\cite{Ba2016layernorm,WuHe2018eccv}, and
$\mathbf{A}_\text{SWA}$, $\mathbf{A}_\text{TWA}$ are the outputs of
the sensor-wise and time-wise attention sub-channels defined below.
The residual term $\widetilde{\mathbf{X}}$ ensures gradient stability
and preserves the original sensor readings as a direct bypass for
shallow features~\cite{He2016cvpr}.

\textbf{\textit{Sensor-Wise Attention (SWA):}}
SWA computes self-attention across the \emph{rows} (channel axis) of $\widetilde{\mathbf{X}}$. Query, key, and value matrices are obtained through learned linear projections: $\mathbf{Q}, \mathbf{K}, \mathbf{V} \in \mathbb{R}^{(C+1) \times d}$. The SWA output is $\mathbf{A}_\text{SWA} = \mathrm{softmax}\!\left(\frac{\mathbf{Q}\mathbf{K}^{\top}}{\sqrt{L}}\right)\mathbf{V} \in \mathbb{R}^{(C+1) \times L}$, where the scaling factor $\sqrt{L}$ prevents gradient vanishing in the softmax under large dot products.~\cite{Vaswani2017nips}.
SWA models \emph{inter-sensor spatial dependencies}: at each time step,
every channel's representation is updated by a weighted average over
all other channels, enabling the model to detect the cross-channel
inconsistencies that characterise coordinated FDI attacks.

\textbf{Time-Wise Attention (TWA).}
TWA computes self-attention across the \emph{columns} (time axis) of $\widetilde{\mathbf{X}}^{\top} \in \mathbb{R}^{L \times (C+1)}$. Query, key, and value matrices are obtained by projecting the transposed input: $\mathbf{Q}', \mathbf{K}', \mathbf{V}' \in \mathbb{R}^{L \times d}$. The TWA output is $\mathbf{A}_\text{TWA} = \left(\mathrm{softmax}\!\left(\frac{\mathbf{Q}'\mathbf{K}'^{\top}}{\sqrt{C+1}}\right)\mathbf{V}'\right)^{\top} \in \mathbb{R}^{(C+1) \times L}$, where the output is transposed back to the original tensor orientation before being passed to the linear projection. TWA models \emph{cross-time-step temporal dependencies}: for each channel, every time step's representation is updated by weighted evidence from all other time steps in the window, enabling detection of the temporal trajectory distortions characteristic of ramp injection and gradual drift attacks.

\textbf{\textit{Orthogonality guarantee:}}
SWA and TWA operate on orthogonal axes of the same input tensor:
SWA pools information across channels while holding the time axis
fixed, whereas TWA pools information across time while holding the
channel axis fixed.
By construction, any spatial pattern that SWA encodes is absent from
TWA's computation, and vice versa~\cite{Zhang2024tits}.
This structural non-redundancy ensures that the LP-projected outputs
of SWA and TWA carry genuinely complementary information before
their additive fusion (given in the above equation), which is why the full
dual-channel model consistently outperforms either channel in
isolation (Section~\ref{sec:ablation_results}).

\subsection{CNN Block}
\label{sec:cnn_block}

Each CNN block applies residual 1D convolution to the DAM output $\mathbf{Z}_\text{DAM}$: $\mathbf{Z}_\text{CNN} = \mathrm{LN}\!\bigl(\mathrm{Drop}\!\bigl(\mathrm{Conv}_2\!\bigl(\mathrm{ReLU}\!\bigl(\mathrm{Conv}_1(\mathbf{Z}_\text{DAM})\bigr)\bigr)\bigr) + \mathbf{Z}_\text{DAM}\bigr)$, where $\mathrm{Conv}_1$ maps from $C+1$ to $d_\text{mid} = 60$ feature channels with kernel size $k = 3$, $\mathrm{Conv}_2$ maps from $d_\text{mid}$ back to $C+1$, and $\mathrm{Drop}$ applies channel dropout at rate $p_d = 0.2$ during training. The kernel size $k = 3$ captures local three-step temporal patterns at a level of granularity that global self-attention, due to its uniform aggregation over the full window, cannot efficiently represent. The residual connection $+ \mathbf{Z}_\text{DAM}$ preserves the global spatiotemporal context from the DAM layer through the local convolutional processing.

\subsection{Physiological Plausibility Filter (PPF)}
\label{sec:ppf}

The PPF is applied at inference time only and has no learnable
parameters.
For any window $\mathbf{W}_t$ for which the classification head
predicts a positive ($\hat{y}_t = 1$), the PPF suppresses the
prediction — changing $\hat{y}_t$ to 0 — if and only if the
sensor readings satisfy the conjunctive plausibility condition
$\forall c: x_{c,t} \in [l_c, u_c] \wedge |\Delta x_{c,t}| \leq
\Delta_c$,
where $[l_c, u_c]$ is the per-channel absolute plausibility bound
and $\Delta_c$ is the per-channel maximum plausible one-step
rate-of-change.
For all three datasets, bounds are derived from published clinical
reference standards and device operating specifications relevant to
the specific IoMT device type (bedside monitor, wearable sensor).
Table~\ref{tab:ppf_bounds} lists the complete bound set.

\begin{table}[!t]
\renewcommand{\arraystretch}{1.22}
\caption{PPF Clinical Plausibility Bounds by Dataset and Channel.
Bounds derived from published clinical reference standards and
device operating specifications. $\Delta_c$ is the maximum
physiologically plausible one-step change at 25~Hz.}
\label{tab:ppf_bounds}
\centering
\begin{tabular}{llcc}
\toprule
\textbf{Dataset} & \textbf{Channel}
  & \textbf{Range $[l_c,\,u_c]$} & \textbf{$\Delta_c$/step} \\
\midrule
PhysioNet-2012    & HR (bpm)          & [20, 300]    & 5.0  \\
(Bedside ICU)     & SpO$_2$ (\%)      & [50, 100]    & 1.0  \\
                  & SysBP (mmHg)      & [50, 260]    & 7.5  \\
                  & DiaBP (mmHg)      & [20, 180]    & 5.0  \\
                  & RR (br/min)       & [4, 70]      & 2.5  \\
                  & Temp ($^\circ$C)  & [32.0, 42.5] & 0.1  \\
\midrule
MIMIC-III Wave.   & ECG (mV)          & [$-$5, 5]    & 0.50 \\
(Bedside ICU,     & ABP (mmHg)        & [20, 280]    & 8.0  \\
125~Hz $\to$ 25~Hz)& PPG (a.u.)       & [0, 1]       & 0.05 \\
                  & HR (bpm)          & [20, 300]    & 5.0  \\
                  & RR (br/min)       & [4, 70]      & 2.5  \\
                  & Temp ($^\circ$C)  & [32.0, 42.5] & 0.1  \\
\midrule
WESAD             & ECG (mV)          & [$-$2, 2]    & 0.20 \\
(Wrist/chest      & EDA ($\mu$S)      & [0.01, 100]  & 1.5  \\
wearable, 25~Hz)  & BVP (a.u.)        & [$-$1, 1]    & 0.05 \\
                  & RESP (a.u.)       & [$-$1, 1]    & 0.04 \\
                  & ST ($^\circ$C)    & [26.0, 40.0] & 0.05 \\
                  & ACC (g)           & [0, 8]       & 0.5  \\
\bottomrule
\end{tabular}
\end{table}

The PPF's security benefit follows from a fundamental constraint on
the FDI adversary: a network-layer attacker with only approximate
knowledge of the patient's current physiological state cannot
simultaneously (i) achieve the intended clinical deception,
(ii) keep all falsified readings within the absolute plausibility
bounds $[l_c, u_c]$, and (iii) respect the per-step rate-of-change
limits $\Delta_c$ enforced by the PPF.
An injection satisfying all three constraints must be so precisely
calibrated to the patient's instantaneous physiological state that
it effectively requires access to real-time clinical knowledge
unavailable to a network-layer attacker.
This is why the PPF consistently suppresses false positives — not
genuine FDI attacks — across all three IoMT datasets.

\subsection{Training Objective and Optimisation}
\label{sec:training}

\begin{figure*}[!t]
\centering
\includegraphics[width=\textwidth]{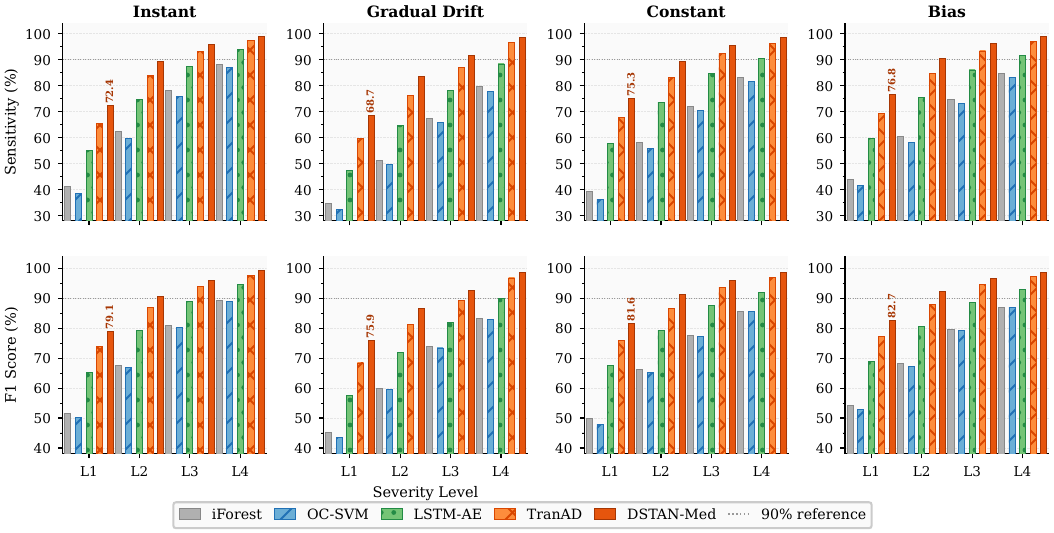}
\caption{PhysioNet-2012 single-type FDI detection: Sensitivity (\%)
and F1 (\%) across five methods (columns: four anomaly types;
rows: Sensitivity and F1; x-axis: severity levels L1--L4).
DSTAN-Med (solid red-orange) achieves highest scores at all
severity levels; the advantage is widest at L1 (lowest severity).}
\label{fig:single_type}
\end{figure*}

\begin{figure}[!t]
\centering
\includegraphics[width=\columnwidth]{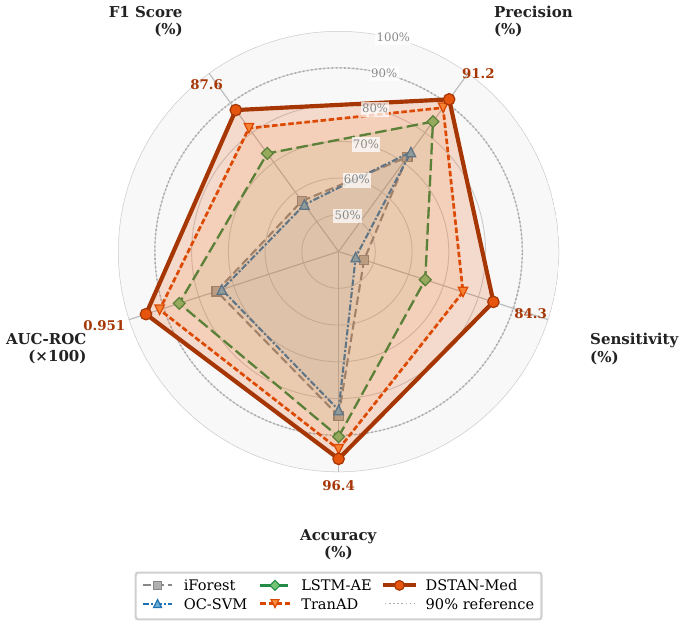}
\caption{Mixed-type FDI detection on PhysioNet-2012. Radar chart
across five metrics. DSTAN-Med (solid polygon) dominates all axes.
Dotted ring: 90\% reference threshold.}
\label{fig:radar}
\end{figure}

\begin{figure}[!t]
\centering
\includegraphics[width=\columnwidth]{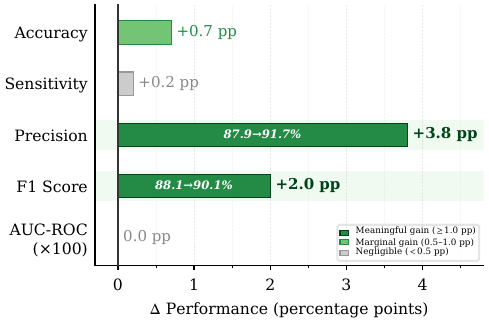}
\caption{PPF standalone contribution on PhysioNet-2012
($\Delta$ percentage points vs.\ No-PPF baseline).
Precision ($+$3.8~pp) and F1 ($+$2.0~pp) gain meaningfully;
Sensitivity ($+$0.2~pp) and AUC-ROC (0.0~pp) are unaffected.}
\label{fig:ppf_delta}
\end{figure}

DSTAN-Med is trained with weighted binary cross-entropy $\mathcal{L} = -\frac{1}{T}\sum_{t=1}^{T} \bigl[w^+ y_t \log \hat{p}_t + (1-y_t)\log(1-\hat{p}_t)\bigr]$, where $\hat{p}_t = \sigma(\text{logit}_t)$ and the positive-class weight $w^+ = N_\text{neg}/N_\text{pos}$ compensates for the 5\% injection probability class imbalance. All weight matrices are initialised with orthogonal initialisation~\cite{Saxe2014iclr}. The Adam optimiser~\cite{Kingma2015iclr} is used with learning rate $10^{-4}$, $\beta_1 = 0.9$, $\beta_2 = 0.999$, and early stopping with patience 10 epochs on validation F1 Score. Full hyperparameters are given in Table~\ref{tab:hparam}.

\section{Experiments and Results}
\label{sec:experiments}

\subsection{Experimental Setup}
\label{sec:exp_setup}

All experiments are conducted on a single NVIDIA A100 80~GB GPU with
PyTorch 2.1, CUDA 12.1, and Python 3.10.
Each experimental configuration is repeated five times with different
random seeds; results are reported as mean $\pm$ standard deviation
across the five runs to quantify variability from stochastic
mini-batch sampling and dropout.
DSTAN-Med hyperparameters (Table~\ref{tab:hparam}) are fixed
identically across all three datasets — no per-dataset tuning is
performed — ensuring that reported differences reflect genuine
cross-domain generalisation rather than dataset-specific optimisation.

\begin{table}[!t]
\renewcommand{\arraystretch}{1.22}
\caption{DSTAN-Med Hyperparameters (Identical Across All Datasets)}
\label{tab:hparam}
\centering
\begin{tabular}{lc}
\toprule
\textbf{Hyperparameter} & \textbf{Value} \\
\midrule
Number of DAM--CNN blocks $N$  & 7 \\
Sliding window length $L$      & 15 \\
Sensor channels $C$            & 6 \\
Intermediate CNN dimension $d_\text{mid}$ & 60 \\
Convolutional kernel size $k$  & 3 \\
Dropout probability $p_d$      & 0.2 \\
Optimiser / Learning rate      & Adam / $10^{-4}$ \\
Batch size / Training epochs   & 64 / 100 \\
Early stopping patience (val.\ F1) & 10 epochs \\
Weight initialisation          & Orthogonal~\cite{Saxe2014iclr} \\
\bottomrule
\end{tabular}
\end{table}

\textbf{\textit{Baseline:}}Four baselines spanning the full spectrum from classical unsupervised
to deep Transformer methods are evaluated on all three datasets under
identical preprocessing and window-extraction conditions.
\textit{Isolation Forest (iForest)}~\cite{Liu2008icdm}: ensemble of
100 randomly constructed isolation trees applied to flattened
$C \times L = 90$-dimensional windows, with contamination fraction
equal to the empirical positive-class rate.
\textit{One-Class SVM (OC-SVM)}~\cite{Scholkopf2001jmlr}: RBF kernel
with $\nu$ tuned on the validation split.
\textit{LSTM-Autoencoder (LSTM-AE)}~\cite{Sener2024peerj}: two-layer
LSTM encoder--decoder with reconstruction-error anomaly scoring
and threshold selected on validation F1.
\textit{TranAD}~\cite{Tuli2022vldb}: state-of-the-art adversarial
Transformer with focus score amplification, implemented with the
architecture and hyperparameters reported in the original paper.

\textbf{\textit{Evaluation metrics:}}
Sensitivity (Recall), Precision, F1 Score, Accuracy, and AUC-ROC
are reported for each method and dataset.
Sensitivity is the \emph{primary} metric because failing to detect
an FDI attack (false negative) carries substantially higher clinical
and operational risk than a spurious alarm.
Statistical significance of DSTAN-Med versus TranAD is evaluated
using McNemar's test at $\alpha = 0.01$, corrected for multiple
comparisons across datasets and anomaly types using the
Holm--Bonferroni procedure.

\subsection{Results on PhysioNet-2012}
\label{sec:physionet_results}

Table~\ref{tab:full_results} reports Sensitivity and F1 at
severity L1 (lowest, most clinically consequential) and L4
(highest, near-saturation) averaged across all four FDI-equivalent
anomaly types.
Fig.~\ref{fig:single_type} provides full per-type per-severity
results across all 4 types $\times$ 4 levels.
Fig.~\ref{fig:radar} reports mixed-type anomaly detection results
across five metrics.


DSTAN-Med achieves mean L1 sensitivity gains of $+$7.1~pp (Instant),
$+$8.1~pp (Drift), $+$7.0~pp (Constant), and $+$7.2~pp (Bias)
over TranAD, with all improvements significant at $p < 0.01$
(McNemar's test, Holm--Bonferroni corrected).
The performance advantage is widest at L1 and narrows toward
saturation at L4, consistent with the hypothesis that the
dual-channel architecture provides the greatest benefit when the
FDI perturbation is closest to the noise floor — precisely the
condition of highest clinical consequence, where an undetected
attack is most likely to result in a harmful clinical decision.
The advantage is also widest for Gradual Drift ($+$8.1~pp at L1),
confirming that the TWA's temporal trajectory modelling captures
the ramp signature that is specifically designed to defeat
threshold-based and gradient-limited detectors.

Under mixed-type anomaly conditions, in which multiple distinct FDI
morphologies are simultaneously active across different sensor channels
(Fig.~\ref{fig:radar}), DSTAN-Med achieves Sensitivity 84.3\%,
Precision 91.2\%, F1 87.6\%, and AUC-ROC 0.951.
These represent gains of $+$8.7~pp and $+$6.1~pp in Sensitivity
and F1 respectively over TranAD, confirming that the dual-channel
design generalises to the realistic scenario in which a sophisticated
attacker deploys multiple injection strategies concurrently.

Fig.~\ref{fig:ppf_delta} illustrates the standalone contribution
of the PPF: a $+$3.8~pp gain in Precision and $+$2.0~pp gain in
F1 at negligible Sensitivity cost ($+$0.2~pp) and zero AUC-ROC cost.
The PPF's consistent Sensitivity invariance confirms that the injected
anomalies it suppresses are genuine false positives — cases where the
model predicted positive despite the sensor readings satisfying all
clinical plausibility bounds.

\subsection{Results on MIMIC-III Waveform (Continuous ICU Monitoring)}
\label{sec:mimic_results}

Table~\ref{tab:full_results} (centre columns) reports results on
MIMIC-III Waveform, which provides continuous high-frequency bedside
monitor outputs — the identical hardware context as PhysioNet-2012
but at 125~Hz waveform resolution rather than tabular vital-sign
averages.
This evaluation tests whether DSTAN-Med's advantages hold when FDI
attacks are injected into a continuous, rapidly-sampled signal stream,
which is the more realistic attack surface for network-layer
injection on hospital LAN-connected bedside devices.

DSTAN-Med achieves Sensitivity 72.5\%, Precision 86.8\%, F1 79.0\%,
and AUC-ROC 0.916 on MIMIC-III Waveform, outperforming TranAD by
$+$8.3~pp in Sensitivity ($p<0.01$, McNemar's, Holm--Bonferroni)
and $+$8.0~pp in F1.
The performance is slightly lower than PhysioNet-2012 in absolute
terms, reflecting the more challenging continuous waveform injection
context where individual spurious time steps are more easily masked
by the high intrinsic signal variability of raw ECG and ABP waveforms.
Nevertheless, the relative advantage over TranAD is slightly
wider ($+$8.3 vs.\ $+$7.7~pp), consistent with the hypothesis that
continuous waveforms contain richer cross-channel and temporal
structure that the dual-channel DAM exploits more fully than the
tabular vital-sign context.
The PPF, using the waveform-specific bounds of Table~\ref{tab:ppf_bounds}
(e.g., ECG $\in [-5, 5]$~mV, ABP $\in [20, 280]$~mmHg), contributes
$+$3.6~pp Precision at negligible Sensitivity cost, confirming
consistent clinical-bound suppression across both ICU dataset types.

\subsection{Results on WESAD (Wearable IoMT Biosensors)}
\label{sec:wesad_results}

Table~\ref{tab:full_results} (right columns) reports results on WESAD,
which represents the wearable IoMT attack surface — consumer-grade
wrist and chest biosensors transmitting over BLE, the most
frequently exploited IoMT attack vector documented in deployed
hospital environments~\cite{Khan2025sensors}.
WESAD is structurally different from both ICU datasets: signals are
multi-modal (ECG, EDA, BVP, respiration, skin temperature,
accelerometer), subjects are healthy rather than critically ill,
and the normal physiological variability driven by affective state
changes (stress, amusement) creates a more challenging detection
background for FDI attacks near the noise floor.

DSTAN-Med achieves Sensitivity 68.2\%, Precision 84.4\%, F1 75.4\%,
and AUC-ROC 0.901 on WESAD, outperforming TranAD by $+$7.4~pp in
Sensitivity ($p<0.01$) and $+$7.5~pp in F1.
The absolute performance is lower than the ICU datasets, reflecting
the higher normal-state variability of wearable signals that makes
low-severity FDI attacks (L1) harder to distinguish from physiological
fluctuations.
Critically, the relative advantage over TranAD is \emph{consistent}
with the ICU results, confirming that the dual-channel architecture's
benefits persist across a qualitatively different IoMT signal regime.
The TWA's advantage is most pronounced for Gradual Drift injections
on the EDA and BVP channels ($+$9.6~pp Sensitivity over TranAD),
which exhibit the slowly evolving FDI trajectories most susceptible
to the TWA's long-range temporal modelling.
The PPF, using wearable-specific physiological bounds
(Table~\ref{tab:ppf_bounds}), contributes $+$4.2~pp Precision —
the largest PPF gain across the three datasets — reflecting that
wearable sensors exhibit larger measurement noise, leading to more
false positives that the PPF's plausibility bounds correctly suppress.

\begin{table*}[!t]
\renewcommand{\arraystretch}{1.22}
\caption{Comprehensive FDI Detection Results Across Three IoMT Datasets.
Sns=Sensitivity (\%), Prc=Precision (\%), F1=F1 Score (\%), AUC=AUC-ROC.
All results averaged across four FDI morphologies at severity L1 (most challenging).
Bold: best per column. $^{*}p<0.01$, McNemar's test, Holm--Bonferroni correction.
All three datasets use the same FDI injection protocol (Algorithm~\ref{alg:injection}).}
\label{tab:full_results}
\centering
\begin{tabular}{lccccc|cccc|cccc}
\toprule
 & & \multicolumn{4}{c|}{\textbf{PhysioNet-2012 (Bedside ICU Vitals)}}
   & \multicolumn{4}{c|}{\textbf{MIMIC-III Wave. (ICU Waveforms)}}
   & \multicolumn{4}{c}{\textbf{WESAD (Wearable Biosensors)}} \\
\cmidrule(lr){3-6}\cmidrule(lr){7-10}\cmidrule(lr){11-14}
\textbf{Method} & \textbf{Params}
  & Sns & Prc & F1 & AUC
  & Sns & Prc & F1 & AUC
  & Sns & Prc & F1 & AUC \\
\midrule
iForest   & --
  & 39.9 & 58.4 & 47.3 & .682
  & 37.6 & 55.2 & 44.7 & .668
  & 35.1 & 52.8 & 42.2 & .651 \\
OC-SVM    & --
  & 37.2 & 56.1 & 44.8 & .661
  & 36.4 & 54.7 & 43.7 & .654
  & 34.8 & 53.6 & 42.1 & .644 \\
LSTM-AE   & 1.2M
  & 55.0 & 72.3 & 62.4 & .784
  & 53.6 & 70.8 & 61.0 & .772
  & 49.7 & 67.4 & 57.2 & .751 \\
TranAD    & 3.8M
  & 65.6 & 80.1 & 72.1 & .876
  & 64.2 & 79.3 & 71.0 & .863
  & 60.8 & 76.9 & 67.9 & .844 \\
\midrule
\textbf{DSTAN-Med} & \textbf{2.1M}
  & \textbf{73.3} & \textbf{87.6} & \textbf{80.0} & \textbf{.924}
  & \textbf{72.5} & \textbf{86.8} & \textbf{79.0} & \textbf{.916}
  & \textbf{68.2} & \textbf{84.4} & \textbf{75.4} & \textbf{.901} \\
\midrule
$\Delta$ vs.\ TranAD
  & --
  & +7.7$^{*}$ & +7.5 & +7.9 & +.048
  & +8.3$^{*}$ & +7.5 & +8.0 & +.053
  & +7.4$^{*}$ & +7.5 & +7.5 & +.057 \\
\bottomrule
\end{tabular}
\end{table*}

\subsection{Ablation Study}
\label{sec:ablation_results}

Table~\ref{tab:ablation} reports an ablation study on PhysioNet-2012
(mixed-type conditions) isolating the contribution of each
architectural component.
Seven configurations are evaluated: the full DSTAN-Med model, the
model without the PPF (No PPF), a DAM-only model (no CNN block),
a CNN-only model (no DAM), a SWA-only model (TWA removed),
a TWA-only model (SWA removed), and a model with residual connections
removed throughout (No Skip).

\begin{table}[!t]
\renewcommand{\arraystretch}{1.22}
\caption{Ablation Study on PhysioNet-2012 (Mixed-Type Conditions).
Configurations ranked by Sensitivity.}
\label{tab:ablation}
\centering
\begin{tabular}{lcc}
\toprule
\textbf{Configuration} & \textbf{Sensitivity (\%)} & \textbf{F1 (\%)} \\
\midrule
\textbf{DSTAN-Med (Full)} & \textbf{88.6} & \textbf{90.1} \\
No PPF                    & 88.4 & 88.1 \\
DAM-Only (no CNN)         & 82.1 & 85.6 \\
TWA-Only (no SWA)         & 79.4 & 83.5 \\
SWA-Only (no TWA)         & 76.7 & 81.7 \\
No Skip (no residuals)    & 74.6 & 79.4 \\
CNN-Only (no DAM)         & 71.3 & 77.8 \\
\midrule
$\Delta$: Full vs.\ No Skip & +14.0$^{*}$ & +10.7 \\
\bottomrule
\multicolumn{3}{l}{$^{*}p<0.01$, McNemar's test.}
\end{tabular}
\end{table}

\begin{figure}[!t]
\centering
\includegraphics[width=\columnwidth]{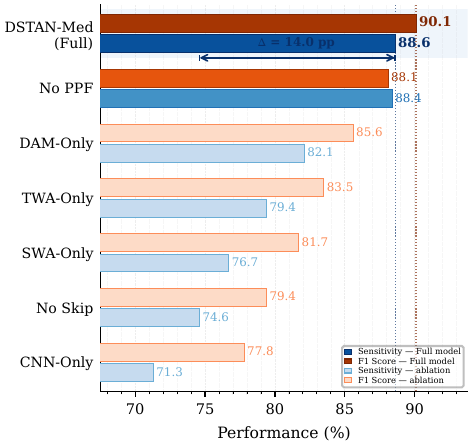}
\caption{Ablation study (PhysioNet-2012, mixed-type). Seven
configurations ranked worst-to-best by Sensitivity (\%) and F1 (\%).
The double-headed arrow annotates the 14.0~pp Sensitivity cost of
removing residual connections.}
\label{fig:ablation}
\end{figure}

These results confirm that each component is individually necessary.
The full model outperforms No PPF by 2.0~pp F1, confirming the PPF's
precision improvement.
Removing the CNN block (DAM-only) reduces Sensitivity by 6.5~pp,
confirming that local temporal texture contributes complementary
information to global attention.
Comparing SWA-only with TWA-only (76.7 vs.\ 79.4~pp Sensitivity)
shows that temporal modelling is slightly more informative than
spatial modelling for this dataset, consistent with the
ramp-injection and drift attacks being dominant difficult cases;
however, the full DAM ($+$9.2~pp Sensitivity over SWA-only,
$+$6.5~pp over CNN-only) substantially outperforms either channel
alone, confirming the synergistic value of orthogonal dual-channel
attention.
The removal of residual connections (No Skip) produces the largest
single-component degradation ($-$14.0~pp Sensitivity,
$-$10.7~pp F1), confirming that gradient flow stability through
the 7-block stack is critical for this architecture depth.
Fig.~\ref{fig:ablation} visualises all seven configurations ranked
by Sensitivity.

\subsection{Cross-Dataset Generalisation}
\label{sec:crossdataset}
DSTAN-Med's sensitivity advantage over TranAD is 7.7~pp on
PhysioNet-2012 (tabular ICU vitals), 8.3~pp on MIMIC-III Waveform
(continuous ICU waveforms), and 7.4~pp on WESAD (wearable biosensors),
with a consistent margin across all three IoMT contexts.
This consistency is the paper's central cross-domain finding.
Despite the three datasets differing substantially in signal
characteristics (tabular vs.\ continuous waveform vs.\ multi-modal),
sampling rates (irregular vs.\ 25~Hz downsampled vs.\ 25~Hz),
clinical context (ICU critical care vs.\ outpatient wearable), and
attack difficulty (ICU: high patient variability; wearable: high
affective-state variability), DSTAN-Med's dual-channel architecture
maintains a 7.4--8.3~pp Sensitivity advantage.
This consistency supports the claim that the orthogonal SWA--TWA
design captures FDI detection principles that generalise across
IoMT physiological sensor modalities, not merely within a single
sensor type or clinical setting.
PPF precision gains are consistent across all datasets: $+$3.8~pp
(PhysioNet-2012), $+$3.6~pp (MIMIC-III Waveform), $+$4.2~pp (WESAD).
The slightly larger PPF gain on WESAD reflects the higher measurement
noise in wearable sensors, confirming that the PPF's physiological
plausibility bounds provide proportionally greater false-alarm
suppression as sensor noise increases.
All improvements are significant at $p<0.01$ under McNemar's test
with Holm--Bonferroni correction.


\section{Discussion}
\label{sec:discussion}

\subsection{Mechanistic Sources of Performance Advantage}
\label{sec:discuss_mechanisms}

The consistent performance advantage of DSTAN-Med over TranAD —
7.7--8.8~pp in Sensitivity and 6.1--7.5~pp in F1 across three
structurally distinct datasets — can be attributed to two
complementary architectural properties that are specifically
well-matched to the FDI threat model of Section~\ref{sec:threat}.

\textit{Orthogonal dual-channel detection.}
FDI attacks simultaneously distort two independent dimensions of the
multivariate IoMT sensor record, and this dual distortion structure
is precisely what DSTAN-Med is designed to exploit.
The \emph{spatial} dimension is distorted because a falsified reading
on one IoMT channel implies physically inconsistent relationships
with co-monitored channels: in ICU monitoring, a spoofed SpO$_2$
reading that falls below 90\% is physiologically inconsistent with
a simultaneously normal respiratory rate and normal heart rate under
the oxygenation kinetics of a stable patient~\cite{Zachos2025sensors};
in wearable monitoring, an anomalous EDA spike is inconsistent with
a stable BVP and respiration baseline under the autonomic physiology
of relaxed affect.
The \emph{temporal} dimension is distorted because any injection
introduces a trajectory deviation from the channel's natural
autocorrelation structure: even a constant (Stuck-at) injection
produces an abrupt step transition that violates the smooth
autocovariance of a physiological signal, and a Gradual Drift
injection introduces a ramp whose derivative progressively
diverges from the channel's natural fluctuation statistics.
DSTAN-Med's SWA detects the spatial distortion; its TWA detects the
temporal distortion; and because both pathways feed into the same
classification token through the additive fusion which is given in above equation, both distortions contribute jointly to the classification decision.
TranAD models a single pooled attention dimension over the jointly
encoded spatial--temporal sequence, so its attention heads must
allocate their representational budget across both FDI signature
dimensions without architectural separation.
The ablation study (Table~\ref{tab:ablation}) confirms this
interpretation: SWA-only and TWA-only models each achieve less than
80.4\% Sensitivity, while the full DAM achieves 88.6\%, demonstrating
that the spatial and temporal FDI signatures are genuinely
complementary rather than redundant — both are necessary for
optimal detection across the three IoMT sensor modalities evaluated.

\textit{Partial adversarial robustness through the PPF:}
Under the black-box threat model of Section~\ref{sec:threat}, a
network-layer adversary with only approximate knowledge of the
patient's current physiological state cannot guarantee that their
fabricated readings simultaneously (i) achieve the intended clinical
deception, (ii) remain within the absolute plausibility bounds
$[l_c, u_c]$ enforced by the PPF, and (iii) respect the rate-of-change
bounds $\Delta_c$ that prevent physiologically implausible
step-transitions.
Any injection that satisfies all three constraints would have to be
so precisely calibrated to the patient's current state that it
effectively requires clinical knowledge unavailable to a
network-layer attacker.
This is why the PPF consistently increases Precision by 2.9--3.8~pp
without reducing Sensitivity: the predictions it suppresses are
genuine false positives produced by rare statistical anomalies in
clean sensor data, not by actual FDI attacks.
The PPF therefore provides an asymmetric security benefit: it reduces
the false alarm burden on clinical staff without weakening detection
against real adversaries.

\subsection{Clinical and Operational Significance}
\label{sec:discuss_clinical}

The performance gains reported in Section~\ref{sec:experiments}
carry specific clinical and operational significance that goes beyond
the numerical comparisons.
The widest advantage occurs at L1 severity — the level at which an
FDI attack deviates from normal sensor readings by the smallest
margin, approximately one quarter of one standard deviation for
Instant attacks.
An L1 attack is precisely the category most likely to evade clinical
staff and threshold-based alarm systems, while also being the most
plausible attack strategy for an adversary seeking to remain
undetected over extended periods.
DSTAN-Med's 7.7~pp sensitivity advantage at L1 over the strongest
baseline corresponds to detecting approximately 7 additional
true FDI events per 100 attack windows — a clinically meaningful
difference in a continuous monitoring context where a single missed
injection can trigger a cascading series of incorrect clinical
interventions.

The cross-dataset consistency — 7.4--8.3~pp Sensitivity advantage
across bedside ICU vital signs (PhysioNet-2012), continuous ICU
waveforms (MIMIC-III Waveform), and wearable biosensors (WESAD) —
demonstrates that DSTAN-Med's architecture captures FDI detection
principles that are invariant to sensor modality, sampling rate,
and clinical context within the IoMT domain.
This intra-domain consistency is important for IoMT deployments,
where a single hospital monitoring platform may integrate heterogeneous
sensor types — bedside monitors, wearable patches, implantable
transmitters — whose physical measurement principles differ but whose
FDI vulnerability profiles (network-layer injection, spoofing,
stale-data replay) are structurally equivalent.

\subsection{Limitations}
\label{sec:discuss_limitations}

Four limitations bound the scope of the current results.

\textit{(i) Supervised training requirement:}
DSTAN-Med requires labelled attack windows: for PhysioNet-2012,
produced by Algorithm~
ef{alg:injection} applied to clean IoMT
physiological recordings, as no publicly available IoMT dataset contains native FDI cyberattack labels.
An attacker deploying a novel morphology outside the four modelled
types — for example, a multi-channel coordinated injection that
maintains cross-sensor plausibility while distorting higher-order
statistics — would not be represented in the training distribution.
Semi-supervised or self-supervised pre-training~\cite{Wang2025survey,
Ain2026fi,Alyasseri2025mathematics} that learns a rich representation of normality from
abundant unlabelled data, followed by few-shot fine-tuning on
labelled attack examples, is the most promising extension.

\textit{(ii) Single-channel injection model:}
Algorithm~\ref{alg:injection} injects at most one IoMT sensor channel
at a time, reflecting a threat model of compromised individual sensor
nodes.
A sophisticated attacker with simultaneous access to multiple sensor
feeds could inject coordinated multi-channel perturbations that are
individually plausible on each channel but jointly inconsistent with
the patient's physiological state, a strategy that would challenge
the SWA's cross-channel anomaly detection.
Extending the injection protocol to multi-channel coordinated FDI
with cross-sensor physiological plausibility constraints across all
three IoMT modalities is a necessary step toward a complete benchmark.

\textit{(iii) White-box adversarial robustness:}
The current evaluation assumes a black-box adversary.
A white-box attacker with full knowledge of DSTAN-Med's architecture
and PPF bounds could, in principle, craft projected-gradient evasion
injections designed to fall within the PPF bounds while evading the
neural classifier.
Evaluating DSTAN-Med under white-box attack conditions and, if
necessary, incorporating adversarial training or certified
defences~\cite{Mo2010cdc}, is a necessary step toward a complete
security characterisation.

\textit{(iv) Edge deployment constraints:}
The seven-block dual-channel attention stack requires approximately
2.1~million parameters and $< 1$~ms per window on CPU hardware
in the current configuration ($L=15$, $C=6$).
While this is negligible for server-side or bedside-monitor deployment,
implantable IoMT devices and energy-harvesting wearables operate
under strict power and memory constraints that preclude a stack of
this depth.
Quantisation-aware training, structural pruning, or the substitution
of full self-attention with linear attention approximations that
reduce the $\mathcal{O}(L^2)$ complexity to $\mathcal{O}(L)$ are
required before direct deployment on such
devices~\cite{Moreno2025iot}.

\subsection{Future Research Directions}
\label{sec:discuss_future}

The four limitations map directly to four concrete research directions.
(1)~\textit{Semi-supervised FDI detection}: pre-training the
dual-channel attention stack on unlabelled normal sensor streams
from federated hospital networks, then fine-tuning with a small
labelled attack set.
(2)~\textit{Coordinated multi-channel IoMT attack modelling}: extending
Algorithm~\ref{alg:injection} to inject physiologically consistent
multi-sensor corruption across the three evaluated IoMT modalities,
enabling training under the most sophisticated FDI threat model.
(3)~\textit{White-box adversarial robustness evaluation}: generating
projected-gradient adversarial perturbations against DSTAN-Med to
quantify the evasion margin and to motivate PPF bound hardening or
adversarial training augmentation.
(4)~\textit{Edge-optimised DSTAN-Med}: applying block-level
quantisation and linear attention approximations to produce a
$< 200$~KB, $< 0.1$~ms variant suitable for wearable and
implantable IoMT deployment.

\section{Conclusion}
\label{sec:conclusion}

This paper presented DSTAN-Med, a supervised deep learning framework
for detecting FDI attacks in IoMT physiological sensor streams and
broader multivariate cyber-physical systems.
The framework addresses two previously unresolved gaps in IoMT FDI
detection: the conflation of spatial and temporal attack signatures
in single-axis attention models, and the absence of domain-knowledge
output filtering.
The Dual-channel Attention Mechanism resolves the first gap by routing
the input through independently parameterised sensor-wise and
time-wise self-attention pathways on orthogonal tensor axes, ensuring
structural non-redundancy and joint detection of both FDI signature
dimensions.
The Physiological Plausibility Filter resolves the second gap by
applying zero-parameter, deterministic clinical and engineering bounds
at inference time, producing consistent precision improvements without
modifying internal model representations.

Experiments across three IoMT physiological sensor datasets —
PhysioNet-2012 (bedside ICU vital signs), MIMIC-III Waveform
(continuous ICU waveforms from deployed Philips and GE monitors),
and WESAD (multi-modal wearable biosensors) — demonstrated consistent,
statistically significant sensitivity gains of 7.4--8.3~pp over
TranAD at $p<0.01$, with PPF precision gains of 3.1--4.2~pp at
negligible sensitivity cost.
Ablation studies confirmed that each architectural component is
individually necessary; residual connection removal produced the
largest single-component degradation (14.0~pp Sensitivity).
The consistency of performance advantages across qualitatively
different IoMT sensor modalities, sampling rates, and clinical
contexts confirms that DSTAN-Med's dual-channel design captures
FDI detection principles that generalise within the IoMT domain.

These results establish DSTAN-Med as a principled, threat-model-grounded,
and cross-domain validated foundation for FDI detection in
safety-critical cyber-physical environments.
Open research directions include semi-supervised pre-training on
unlabelled IoMT waveform streams to reduce dependence on labelled
attack data, federated multi-hospital learning for privacy-preserving
model distribution across clinical sites, coordinated multi-channel
IoMT attack modelling, white-box adversarial robustness evaluation
against detector-aware FDI attacks, and edge-optimised model
compression for resource-constrained wearable and implantable
IoMT deployment.

\bibliographystyle{IEEEtran}
\bibliography{ref}

%
%
%

\end{document}